\documentclass[amsmath,amssymb,aps,prx,reprint,superscriptaddress,longbibliography]{revtex4-2}

\usepackage{amsmath}
\usepackage[dvipsnames,svgnames]{xcolor}
\usepackage{amsfonts,amssymb}
\usepackage{graphicx}
\usepackage{academicons}
\usepackage{multirow}
\usepackage{orcidlink}
\usepackage{booktabs}
\usepackage{wasysym}
\usepackage{needspace}
\usepackage[version=4]{mhchem}
\usepackage{glossaries}
\usepackage[activate={true,nocompatibility},final,tracking=true]{microtype}
\SetTracking{encoding={*}, shape=sc}{40}

\usepackage{fancyhdr}

\makeatletter
\let\ps@plain\ps@fancy
\makeatother

\usepackage{hyperref}
\hypersetup{
pdfnewwindow=true,          
colorlinks=true,            
linkcolor=DarkBlue,         
citecolor=DarkBlue,         
filecolor=DarkBlue,         
urlcolor=DarkBlue           
}
\begin{document}
\title{A Perspective on Training Machine Learning Force Fields for \\Solid-State Electrolyte Materials}

\author{Zihan Yan\,\orcidlink{0000-0002-8911-6549}}
\affiliation{School of Materials Science and Engineering, Zhejiang University, Hangzhou, China}
\affiliation{Department of Materials Science and Engineering, Westlake University, Hangzhou, China}

\author{Shengjie Tang\,\orcidlink{0000-0001-8036-5189}}
\affiliation{School of Materials Science and Engineering, Zhejiang University, Hangzhou, China}
\affiliation{Department of Materials Science and Engineering, Westlake University, Hangzhou, China}

\author{Yizhou Zhu\,\orcidlink{0000-0002-5819-7657}}
\email[]{zhuyizhou@westlake.edu.cn}
\affiliation{Department of Materials Science and Engineering, Westlake University, Hangzhou, China}

\date{\today}

\begin{abstract}

\begin{center}
	\textbf{Abstract}
\end{center}
\vspace{-0.2cm}

	Machine learning force fields enable high-accuracy modeling of solid-state electrolytes (SSEs). This perspective evaluates dataset size, reference quality, and model architectures. We show that rigid SSE frameworks favor efficient learning, prioritizing data quality over quantity. Crucially, force RMSE does not reliably predict transport performance. By analyzing locality and benchmarking frameworks, we provide practical guidelines to accelerate the development of next-generation solid-state batteries.

\end{abstract}

\maketitle

\section{Introduction}

Solid-state electrolytes (SSEs) are ceramic materials with exceptionally high ionic conductivities, which can potentially enable next-generation safer, higher-energy-density solid-state batteries~\cite{zheng2018review,huang2025solid}. Understanding atomic-level ionic diffusion mechanisms are essential for the rational design of SSE materials~\cite{jun2024diffusion}. However, such atomic-level insights sometimes exceed the capabilities of experimental characterization alone. Therefore, computational methods, particularly molecular dynamics (MD) simulations, play an indispensable role in elucidating ion diffusion processes at the atomic scale. Traditional approaches face inherent limitations: \textit{Ab initio} molecular dynamics (AIMD) provides quantum-mechanical accuracy but is restricted to small scales and short times, often necessitating elevated temperatures for statistical convergence. Conversely, classical MD with empirical potentials offers efficiency but often lacks the accuracy and transferability required for complex materials, risking underfitting due to their rigid functional forms.

In contrast, machine learning molecular dynamics (MLMD) simulations have emerged as a powerful approach that offers \textit{ab initio} accuracy at computational costs comparable to classical potentials. MLMD simulations rely on machine learning force fields (MLFFs), which utilize a flexible, data-driven approach to capture complex interatomic interactions. Notable MLFF frameworks include gaussian approximation potential~\cite{bartok2010gaussian}, moment tensor potential~\cite{shapeev2016moment,novikov2020mlip}, deep potential~\cite{zhang2020dp,zeng2023deepmd}, neuroevolution potential (NEP)~\cite{fan2021neuroevolution,song2024general}, atomic cluster expansion~\cite{drautz2019atomic}, MACE~\cite{Batatia2022Design}, NequIP~\cite{batzner2022}, Allegro~\cite{musaelian2023learning}, CACE~\cite{cheng2024cartesian}, \textit{etc}. For specific material systems, MLFFs are trained on datasets generated from high-accuracy quantum mechanical methods, such as density functional theory (DFT). Once successfully trained, these MLFFs achieve DFT-comparable accuracy, far exceeding empirical force fields while maintaining computational costs orders of magnitude lower than first-principles methods~\cite{unke2021machine}. The combination of high accuracy and low cost makes MLFFs well-suited for atomistic modeling in complex material systems, particularly in MD simulations over extended spatial and temporal scales~\cite{ying2025advances}. Recent MLFF applications in SSEs have demonstrated successes in revealing mechanisms of defect chemistry on ionic conductivities~\cite{yan2024impact}, uncovering order-disorder transition mechanisms~\cite{yan2024impact,wang2023frustration,geng2024elucidating}, and screening composition spaces~\cite{du2025assessment}. Despite this rapid progress, training MLFFs for specific SSEs remains a ``black box" process heavily reliant on individual experience. The unique superionic features of SSEs complicate decisions regarding training set size, reference DFT fidelity, and architecture selection. As typical ionic materials, it remains unclear how long-range Coulomb interactions affect the MLFF performance, since many MLFF frameworks utilize local descriptors. Without systematic benchmarks or general recommendations available for the community, developing reliable MLFFs for SSEs still relies largely on personal expertise.

In this perspective, we share our experience and insights in developing MLFFs for various SSE materials. Our analysis shows that, due to their rigid structure and restricted diffusion pathways, the potential energy surfaces (PESs) of SSEs can be captured by modest-size datasets. This finding challenges the common assumption that large training datasets are always necessary to describe ion diffusion in SSEs. We investigated the critical role of DFT reference data quality and the trade-off between model complexity and efficiency. Our benchmark reveals that force RMSE is not a definitive indicator of model physical reliability, especially for transport properties. With systematic locality tests, we clarified why short-range models remain effective for SSE materials and provided practical guidance for balancing accuracy and scale in ion transport simulations. We hope these recommendations help the community accelerate the development of MLFFs for SSEs and advance study for next-generation solid-state batteries.

\section{How large a dataset is needed to train MLFFs for SSEs?}

\begin{figure*}[]
	\centering
	\includegraphics[width=0.75\linewidth]{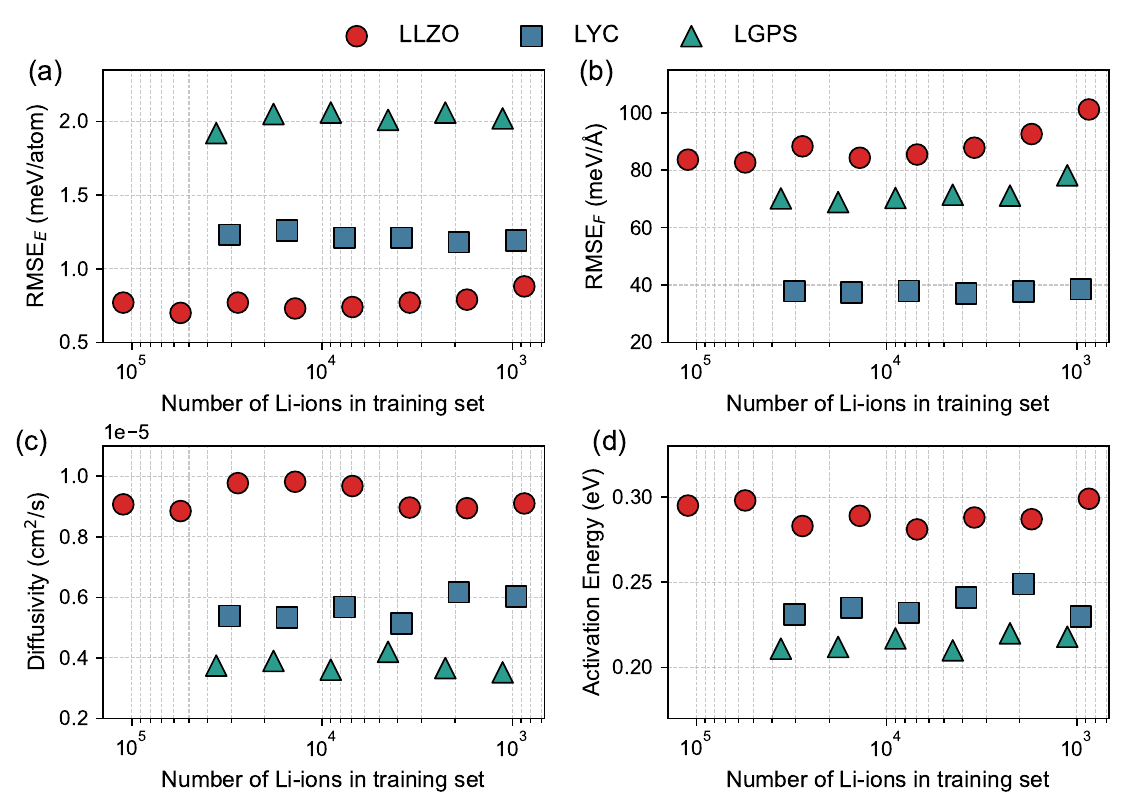}
	\caption{Dataset-size sensitivity tests on MLFF performance in \ce{Li7La3Zr2O_{12}} (LLZO), \ce{Li3YCl6} (LYC), and \ce{Li_{10}GeP2S_{12}} (LGPS). The validation root mean square errors of (a) energy and (b) force. (c) The diffusivities of Li-ions (1000K for LLZO, 500K for LYC and LGPS). (d) The activation energies of Li-ions.}
	\label{fig:li_num}
\end{figure*}

Conventionally, SSEs are regarded as multi-component materials with relatively complex configurations due to the mobile ion sublattice. Therefore, to ensure adequate sampling of the PES, it is widely assumed that a large training dataset, typically consisting of at least a few thousand configurations, is required to develop an accurate MLFF for a SSE material. For instance, studies on the $P\bar{3}m1$ phase of \ce{Li3YCl6} utilized about 1,700--1,850 structures~\cite{zhang2024size,wang2023frustration}, while similar benchmarks across various ternary halides and sulfides consistently employed 1,800 configurations~\cite{xu2023machine}. The dataset size scale increases dramatically when studying complex phenomena. For example, investigations into the crystallization of \ce{Li3PS4} glass and \ce{Li2ZrCl6} transport mechanisms have relied on much larger datasets, containing 34,510 and 60,000 structures, respectively~\cite{zhou2024insights,guo2025unveiling}. It appears that datasets containing thousands to tens of thousands of structures have become the ``standard" for simulating complex ion transport and phase behavior in SSEs. However, several tests presented in this Perspective suggest a different picture: the PES of crystalline SSEs is relatively easy to sample, and therefore can be effectively captured with a small amount of training data.

In fact, the inherent structural characteristics of SSEs determine this ease of sampling. While the mobile ion sublattice renders SSEs more complex than typical rigid solids, their PESs remain significantly simpler than those of liquid, molecular, or polymer systems. In liquid or polymer systems, atoms can explore vast, unconstrained configurational spaces. In contrast, crystalline SSEs possess a rigid framework that provides a stable structural backbone. Within this framework, host atoms only vibrate around equilibrium positions with small displacements, while the mobile ions, despite highly diffusive, are restricted to predefined pathways and specific crystallographic sites. From a sampling perspective, this means SSEs behave more like traditional solids than unconstrained liquids, allowing a relatively small number of representative structures to provide sufficient coverage of the relevant local atomic environments.

To quantify this data efficiency, we conducted systematic dataset-size sensitivity tests on three representative SSE materials: the oxide \ce{Li7La3Zr2O_{12}} (LLZO), the halide \ce{Li3YCl6} (LYC), and the sulfide \ce{Li_{10}GeP2S_{12}} (LGPS). These materials span different chemistries, crystal structures, and local Li-ion environments, enabling a comprehensive assessment of the generality of our observations. For LLZO, we utilized the PBEsol dataset from our previous work~\cite{yan2024impact}, which contains 1,978 LLZO configurations with 56 Li-ions per structure, totaling 110,768 local Li-ion environments. Regarding the MLFF framework, we adopted the fourth-generation NEP model proposed by Fan \textit{et al.}~\cite{song2024general}, which offers a balance between simulation accuracy and speed~\cite{yan2024impact}, and minimize our testing costs. The NEP model achieved sub-meV/atom precision when trained on the full dataset of LLZO: the root mean square errors (RMSEs) for energy and force are 0.77 meV/atom and 83.67 meV/Å, respectively. We then progressively reduced the dataset by factors of $1/2^n$ ($n = 1, 2, ...$) using the farthest-point sampling algorithm implemented in our GPUMDkit package~\cite{yan2025gpumdkit}. Each reduced dataset was used to train a new NEP model, while the full 1,978-structure dataset served as a test set to evaluate how prediction accuracy evolves as the training set size decreases.

Remarkably, even aggressive dataset reduction has a minimal impact on model accuracy (Figure~\ref{fig:li_num}(a) and (b)). For the LLZO system, the RMSE for energy and force remains nearly constant as the dataset is reduced to 1/32 of its original size (\textit{i.e.}, $n=5$, corresponding to only 62 configurations containing 3,472 Li-ions). A slight upward trend in error emerges when the reduction factor reaches $n=6$ and $n=7$. Even at an aggressive reduction factor of $1/128$ ($n=7$), where only 15 LLZO structures (corresponding to 840 Li-ions) are retained, the test errors increase only modestly to 0.88 meV/atom and 101.17 meV/Å. This robust performance demonstrates that fewer than 1,000 local Li-ion environments are sufficient to capture the essential interatomic interactions of LLZO. 

More importantly, we evaluated whether these data-efficient models could maintain an accurate description of ion transport properties, which is a fundamental requirement for MLFF in SSE research. As shown in Fig.~\ref{fig:li_num}(c), the calculated Li-ion diffusivities at 1000 K remain consistent across all models with different training set sizes, with values around 0.88--0.98 $\times$ 10$^{-5}$ cm$^2$/s. Similarly, the activation energies ($E_a$) calculated from the Arrhenius plots at four temperatures (950, 1000, 1100, 1200 K) cluster closely between 0.281 and 0.299 eV (Fig.~\ref{fig:li_num}(d)), which is well within an acceptable range. The complete Arrhenius plots are provided in Fig.~S1 of Supporting Information (SI).

Furthermore, we performed similar dataset-size sensitivity analyses on two chemically distinct SSE materials: the halide LYC and the sulfide LGPS. For LYC, we employed the PBE+optB88-vdW dataset for the P$\bar{3}$m1 phase from Wang \textit{et al.}~\cite{wang2023frustration}, containing 1,698 structures with 18 Li-ions each, totaling 30,564 local Li-ion environments. For LGPS, we utilized part of the PBEsol dataset from Huang \textit{et al.}~\cite{huang2021deep}, containing 1,799 unit-cell structures with 20 Li-ions each, totaling 35,980 local Li-ion environments. As shown in Fig.~\ref{fig:li_num}, both halide and sulfide electrolytes exhibit similar trends as observed in LLZO. Specifically, LYC exhibits insensitivity to dataset reduction, with its energy and force RMSEs remaining almost unchanged even when the number of Li-ions drops below 1,000. While the force RMSEs for LGPS show a slight upward trend as it approaches this threshold. However, the impact on the predicted transport properties remains within acceptable ranges. Across all three chemically diverse SSE families (oxide, halide, and sulfide), approximately 1,000 Li-ion local environments appear sufficient to train a well-performing force fields under the NEP framework. Interestingly, our tests using the MACE framework show that its error is more sensitive to dataset size, with RMSEs exhibiting a more stable increase as the training set decreases (see Fig.~S2 in SI).

We attribute this disparity to the underlying optimization algorithms. While NEP employs the separable natural evolution strategy (SNES)~\cite{schaul2011high}, MACE relies on gradient descent-based training, which typically requires larger datasets to effectively explore the complex parameter spaces. To isolate the impact of the optimization algorithm, we performed tests using GNEP framework~\cite{huang2026efficient}, which employs the same architecture as NEP but uses a gradient descent-based optimizer similar to MACE. As shown in Fig.~S3, GNEP also displays a more pronounced sensitivity to data sparsity compared to the standard NEP. This confirms that the superior data efficiency of the standard NEP is also attributed to its SNES algorithm.

These results challenge the conventional wisdom that accurate simulation of ion diffusion in SSEs requires massive training datasets. In fact, the rigid framework and constrained diffusion paths of Li-ions make its PES relatively easy to learn. Furthermore, once Li-ions begin to diffuse in its channels, they naturally explore most of the relevant local environment and can effectively self-sample the configuration space even in relatively short trajectories. This observation shifts the emphasis of MLFF development for SSEs: choosing structures that can effectively cover the relevant configuration space is far more important than accumulating a large dataset.

\section{What quality of DFT reference data is necessary?}

\begin{figure*}[]
	\centering
	\includegraphics[width=0.95\linewidth]{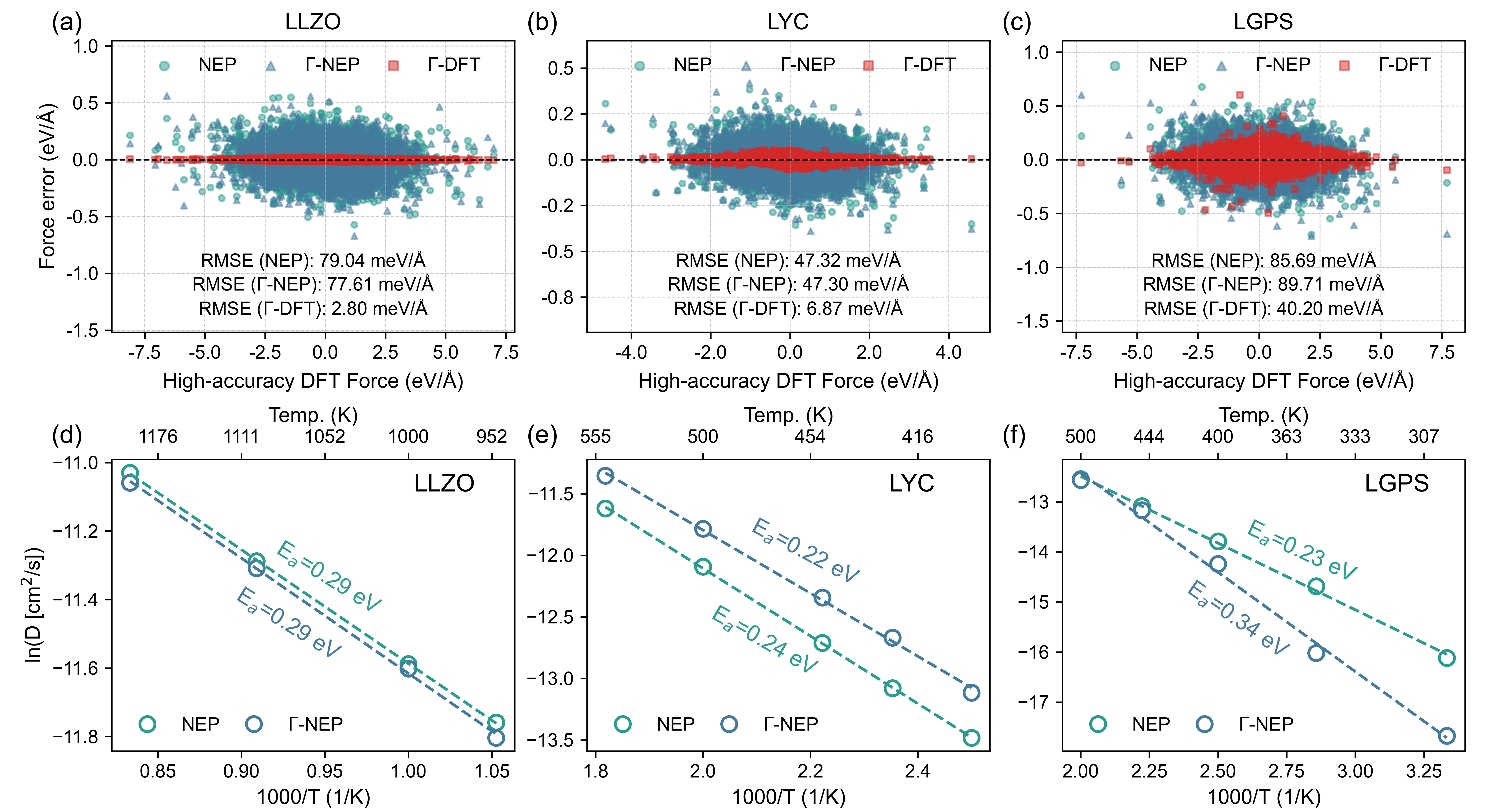}
	\caption{(a-c) Force error distributions for LLZO, LYC, and LGPS comparing $\Gamma$-point DFT ($\Gamma$-DFT) and two NEP models.  NEP is trained on high-accuracy data and $\Gamma$-NEP is trained on $\Gamma$-DFT data. The high-accuracy DFT data are used as reference. (d-f) Arrhenius plots of Li-ions diffusivities calculated by NEP and $\Gamma$-NEP models.}
	\label{fig:force_arrhenius}
\end{figure*}

For efficient and reliable MLFF development, DFT calculation settings are a crucial factor yet receives less attention than dataset size or diversity. Herein, we evaluated how \textit{k}-point density in DFT calculations affects force field performance for LLZO, LYC, and LGPS. We selected 50, 150, and 150 structures from their datasets and performed DFT calculations at two levels of accuracy: high-accuracy calculations with KSPACING = 0.25 Å$^{-1}$, and low-accuracy $\Gamma$-point-only calculations. The high-accuracy setup represents a common setting for self-consistent field (SCF) calculations, whereas the $\Gamma$-point-only approach is routinely used to improve computational efficiency during AIMD simulations. We subsequently trained two sets of NEP models: NEP (on high-accuracy DFT data) and $\Gamma$-NEP (on $\Gamma$-point DFT data) to directly compare how reference data quality impacts model performance.

Figures~\ref{fig:force_arrhenius}(a)-(c) compare the forces calculated using high-accuracy DFT and $\Gamma$-DFT. For LLZO, with its relatively large unit cell ($\sim$12--13 Å in box length, 192 atoms), using only the $\Gamma$-point for DFT calculations basically does not sacrifice or affect the calculation accuracy of its forces, as shown by the red points in Fig.~\ref{fig:force_arrhenius}(a). The force RMSE between $\Gamma$-DFT and high-accuracy DFT is only 2.80 meV/Å. In contrast, $\Gamma$-DFT of LYC and LGPS, with their smaller cells ($\sim$ 9--12 Å with 60 atoms for LYC, $\sim$9--13 Å with 50 atoms for LGPS), show larger force errors of 6.87 and 40.20 meV/Å, respectively. Importantly, all these DFT-level force errors are smaller than the NEP fitting errors. For LLZO, both NEP and $\Gamma$-NEP exhibit force RMSEs around 80 meV/Å, far greater than the $\Gamma$-DFT error (2.80 meV/Å). This suggests that current NEP models remain in an underfitting regime, unable to capture all features present in the training data, and that DFT accuracy is not the limiting factor for reducing force RMSEs.

However, for MLFFs applied to SSE materials, a low RMSE is not the ``gold standard" for model reliability. Instead, we are more interested in whether the MLFF can accurately describe ion transport processes. For LLZO (Fig.~\ref{fig:force_arrhenius}(d)), NEP and $\Gamma$-NEP yield nearly identical diffusivities and activation energies (0.29 eV), which is expected given that both DFT and $\Gamma$-DFT provide sufficiently accurate reference data. For LYC (Fig.~\ref{fig:force_arrhenius}(e)), the two models produce similar activation energies (0.22 \textit{vs.} 0.24 eV), though $\Gamma$-NEP shows a slightly larger prefactor, resulting in systematically higher diffusivities across all temperatures. This deviation, while detectable, remains within acceptable bounds.

LGPS (Fig.~\ref{fig:force_arrhenius}(f)), however, reveals the critical importance of reference data quality. Despite NEP and $\Gamma$-NEP showing nearly identical force RMSEs relative to high-accuracy DFT (85.69 vs. 89.71 meV/Å), their transport predictions diverge dramatically: $\Gamma$-NEP predicts $E_a$ = 0.34 eV, substantially higher than the 0.23 eV obtained with NEP. This overestimation would lead to qualitatively incorrect conclusions about material diffusion properties. This poor performance originates from the $\Gamma$-NEP model being trained on low-accuracy, $\Gamma$-point-only DFT data. In the LGPS system, the relatively small unit cell causes $\Gamma$-point sampling to introduce the largest reference errors (40.20 meV/Å) compared to high-accuracy DFT data. This comparison clearly demonstrates that force RMSE is not a reliable indicator of force field quality for ion transport applications.

To further probe this issue, we trained three independent replicas of both NEP and $\Gamma$-NEP models using different random initializations (see Fig.~S4 in SI). For models trained on high-accuracy data, all three replicas converge to nearly identical transport predictions, demonstrating robust reproducibility. In stark contrast, $\Gamma$-NEP models trained on low-quality data exhibit substantial variation across replicas, with LGPS activation energies ranging from 0.26 to 0.34 eV (Fig.~S4 (c) in SI). This statistical analysis confirms that inadequate reference data quality not only biases predictions but also increases their uncertainty, making model behavior unpredictable. These findings serve as a critical caution for the community: training MLFFs directly on raw AIMD snapshots is inherently risky, as such data often suffer from limited \textit{k}-point density and lower convergence settings. Such a shortcut may lead to models that are not only inaccurate but also statistically unstable. Therefore, we strongly recommend the standard practice of performing additional SCF calculations on extracted snapshots to ensure the reliability of the training set.

While these results suggest that common SCF settings are sufficient for NEP, it raises a critical question: is this ``sufficiency" merely a consequence of the limited expressivity of the NEP framework? In other words, if a more advanced architecture were employed, would the same DFT accuracy still be sufficient? To address this, we conducted additional tests using the state-of-the-art MACE model. Notably, even this highly expressive architecture exhibits a force RMSE of approximately 8 meV/Å for LLZO (Fig.~\ref{fig:rmse-ea}(a)), which is still significantly larger than the error introduced by $\Gamma$-point DFT (2.8 meV/Å). Furthermore, the ion transport properties predicted by the MACE model agrees well with those predicted by the NEP model (Fig.~\ref{fig:rmse-ea}(d)). This indicates that the accuracy of common SCF calculations has surpassed the fitting capabilities of current MLFF frameworks. This is particularly true for short-range, non-message-passing models like NEP. For such model, the common static precision provides a sufficiently accurate baseline, making more precise and expensive settings unnecessary for achieving reliable transport predictions.

These findings lead to clear practical recommendations. First, when generating training data for SSE force fields, prioritize DFT calculation quality over dataset size. High-accuracy calculations on fewer structures will generally outperform low-accuracy calculations on larger datasets. Second, the common SCF setup is precise enough to get the DFT reference data. For practical efficiency, constructing a larger supercell combined with $\Gamma$-point calculations offers an optimal strategy. This approach provides sufficient DFT accuracy while increasing the number of atoms (and thus local atomic environments), as demonstrated by the success of LLZO models. Third, RMSE should not be used as the primary metric for evaluating force field quality. The magnitude of the residual RMSE reflects the inherent limitations of the force field architecture, not the training quality. For example, a consistently high RMSE may indicate that the model fails to capture certain key features, such as long-range interactions, leading to underfitting that cannot be resolved through further training. As we will discuss next, a major source of the limitations of such architectures is the handling of interatomic interactions, especially long-range interactions.

\section{How Long-range Interactions Affect Force Field Accuracy and Performance?}

To evaluate how long-range interactions affect the accuracy and performance of MLFFs, we benchmarked two NEP-based and four MACE-based models. The standard short-range NEP serves as our baseline, while the qNEP model~\cite{fan2026qnep} incorporates the Latent Ewald Summation (LES) method~\cite{cheng2025latent} to explicitly capture Coulomb interactions via dynamically inferring atomic charges from the local atomic environment. For the MACE framework, we examined four variants (MACE-T0-I, MACE-T0-E, MACE-T1-E, and MACE-T1-E-LES) to decouple the effects of model architecture. Here, the suffix T0/T1 denotes zero (local) or one message-passing layer, while I/E distinguishes between invariant and equivariant models.

\begin{figure*}[]
	\centering
	\includegraphics[width=0.95\linewidth]{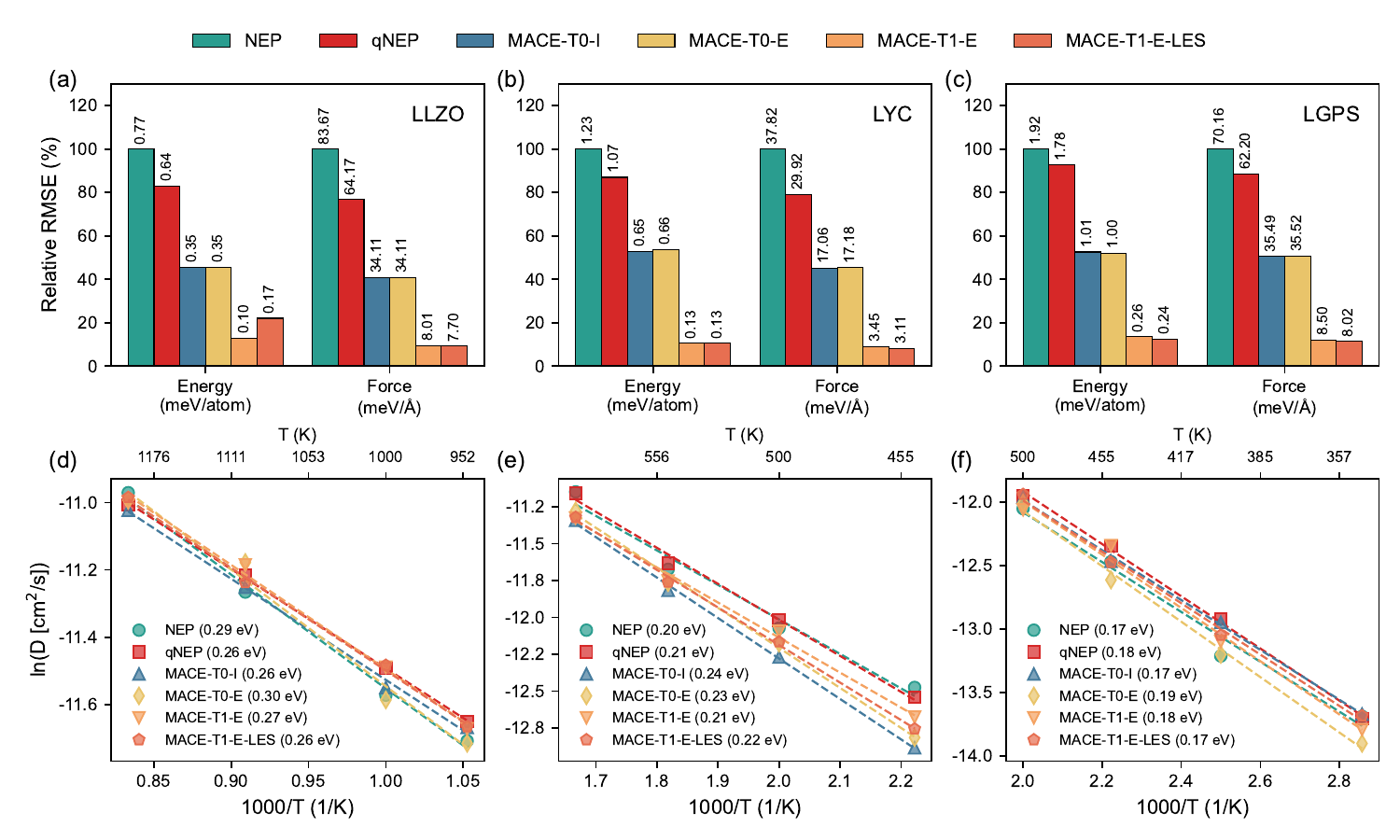}
	\caption{(a-c) Relative RMSE of energies and forces for LLZO, LYC, and LGPS, predicted by different models. (d-f) Arrhenius plots of Li-ions diffusivities for the corresponding materials, calculated using different models.}
	\label{fig:rmse-ea}
\end{figure*}

Figures~\ref{fig:rmse-ea}(a)-(c) present relative RMSEs for energies and forces across the three materials. For LLZO, the baseline NEP model exhibits energy and force errors of 0.77 meV/atom and 83.67 meV/Å, respectively. Though highest among tested models, these errors remain within acceptable ranges for most MD simulations. The qNEP model, incorporating long-range interactions, improves accuracy by reducing 20\% of the RMSE relative to the NEP model. This improvement highlights the benefits from explicit long-range electrostatic treatment in qNEP. Interestingly, even without explicit long-range treatment or message passing, MACE-T0-I and MACE-T0-E exhibit comparable accuracy, with both yielding energy and force RMSEs approximately 50\% lower than the NEP baseline. Since both MACE-T0 models lack message passing and use the same cutoff as the NEP model, their superior accuracy stems primarily from the higher-order expansion of the architecture. The near-identical performance of the invariant (T0-I) and equivariant (T0-E) variants indicates that explicit equivariant features offer limited additional benefit in improving training accuracy.

The lowest error in energy and forces come from the MACE-T1-E and MACE-T1-E-LES models, both incorporate message passing mechanisms. They achieve remarkably low force RMSE values (8.01 and 7.70 meV/Å), approximately 10\% of the NEP baseline (83.67 meV/Å), representing an order-of-magnitude improvement in training accuracy. However, further adding LES to MACE-T1-E provides minor accuracy enhancement, contrasting sharply with $\sim$20\% improvement observed when adding LES to NEP (qNEP). These findings may suggest that the message passing mechanism in MACE-T1-E already establishes a sufficiently extended receptive field to implicitly capture long-range interaction. Therefore, explicit long-range Coulombic terms have limited effects on force field accuracy in these cases.

These accuracy trends are consistently observed across LGPS and LYC (Figures~\ref{fig:rmse-ea}(b) and (c)). Across all three materials, the RMSE of qNEP is approximately 80\% of the NEP baseline, while the errors for MACE-T0 and MACE-T1 models consistently represent approximately 50\% and 10\% of the NEP baseline, respectively. This consistent hierarchy across different chemical systems confirms that the observed accuracy gains are intrinsic to the higher-order expansion, message-passing mechanism and LES approach.

However, as emphasized earlier, energy and force RMSEs alone cannot serve as a definitive criterion for force field quality; verification must come from target properties of interest. Fig.~\ref{fig:rmse-ea}(d)-(f) presents Arrhenius plots of Li-ion diffusivities calculated using each model. For LLZO, despite order-of-magnitude disparities in RMSEs, all models yield relatively consistent diffusion behaviors with activation energies of 0.26--0.30 eV, which are acceptable for transport property predictions. Similar consistency was observed for LYC and LGPS, demonstrating that while MACE models achieve superior RMSE metrics, their predictions of ion transport properties exhibit no significant difference compared to NEP and qNEP models. All results are within acceptable error ranges.

One noteworthy discrepancy appears in LGPS activation energies. The calculated value here 0.18 eV is lower than the aforementioned result 0.23 eV  (Fig.~\ref{fig:force_arrhenius}(f)). This discrepancy arises from the different sizes of simulation cells. Due to the high computational cost of MACE, we adopted small simulation cells to ensure a fair comparison across all models. In this benchmark, the simulation cell of LLZO, LYC and LGPS has 192, 120 and 200 atoms, respectively. LGPS exhibit known finite-size effects that artificially lower the activation energy in smaller supercells, as documented in previous studies~\cite{huang2021deep}. In fact, finite-size effects are not unique to LGPS. Previous reports show that size-dependent diffusion behavior also exhibits in tetragonal LLZO and LYC at low temperatures~\cite{klenk2016finite,zhang2024size}. However, simulations on these materials using larger cells present significant challenges for the MACE models. We therefore omitted such tests in this comparison.

\begin{figure}[]
	\centering
	\includegraphics[width=0.8\linewidth]{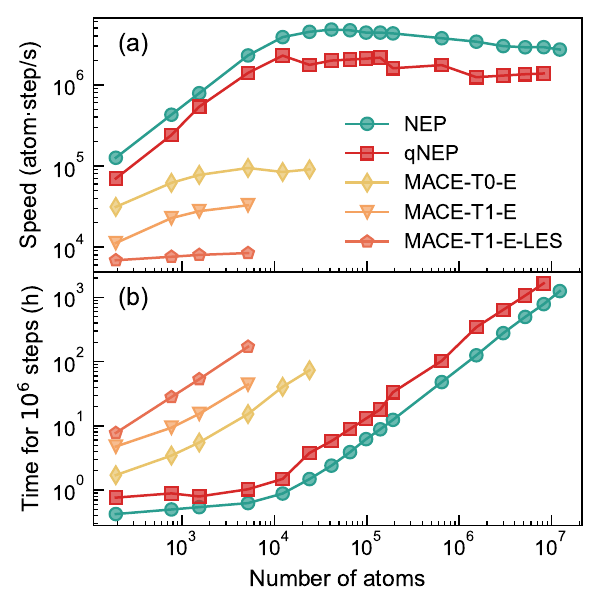}
	\caption{Scaling of (a) simulation speed and (b) total wall-clock time required for $10^6$ MD steps as a function of the number of atoms. All tests were performed on the LLZO system using a single NVIDIA V100 GPU (32 GB).} 
	\label{fig:speed}
\end{figure}

These observations underscore the critical trade-off between accuracy and efficiency. For SSEs, achieving sufficient spatial scale is often as important as force field precision, as large supercells are necessary to mitigate finite-size artifacts and ensure robust statistical sampling. While advanced architectures offer superior training accuracy, their high computational cost can prohibit the large spatiotemporal scales required for reliable transport modeling. Thus, selecting a framework that balances model complexity with scalability deserves careful consideration.

To quantify computational efficiency, we benchmarked all models using the LLZO systems on a single NVIDIA V100 32GB GPU at the Open Source Supercomputing Center of S-A-I. Due to implementation constraints, different MD packages were employed: GPUMD~\cite{xu2025mega} for NEP/qNEP and LAMMPS/ASE~\cite{plimpton1995fast,larsen2017atomic} for MACE (see SI for details). The simulation efficiency differs significantly across these frameworks (Figure~\ref{fig:speed}). NEP and qNEP are one to two orders of magnitude faster than MACE variants, maintaining this advantage across system sizes from hundreds to millions of atoms. Remarkably, a single V100 32GB GPU can handle tens of millions of atoms using NEP or qNEP. In contrast, MACE faces severe memory limitations. Even the simplest variant, MACE-T0-E, is memory-limited to around 10,000 atoms, making it impossible to simulate larger systems on this hardware. The simulation speed of MACE-T0-I is almost the same as that of MACE-T0-E, so its performance is not listed separately.

Fig.~\ref{fig:speed}(b) illustrates wall-clock time required for one million MD steps, a common production run length for diffusivity calculations in MD simulations. For LLZO systems with fewer than 10,000 atoms, NEP and qNEP can complete such simulations within an hour, while the MACE model requires several or even tens of hours. Despite the RMSE of MACE is reduced by an order of magnitude, the calculated diffusion properties are not significantly different from results from NEP (Fig.~\ref{fig:rmse-ea}). Therefore, GPUMD and NEP achieve an excellent balance between accuracy and efficiency, making them particularly suitable for scenarios requiring large-scale, long-time simulations, such as ion transport behavior in SSE materials.

\section{How much impact does including long-range interactions in MLFF actually have on SSE?}

\begin{figure*}[]
	\centering
	\includegraphics[width=0.95\linewidth]{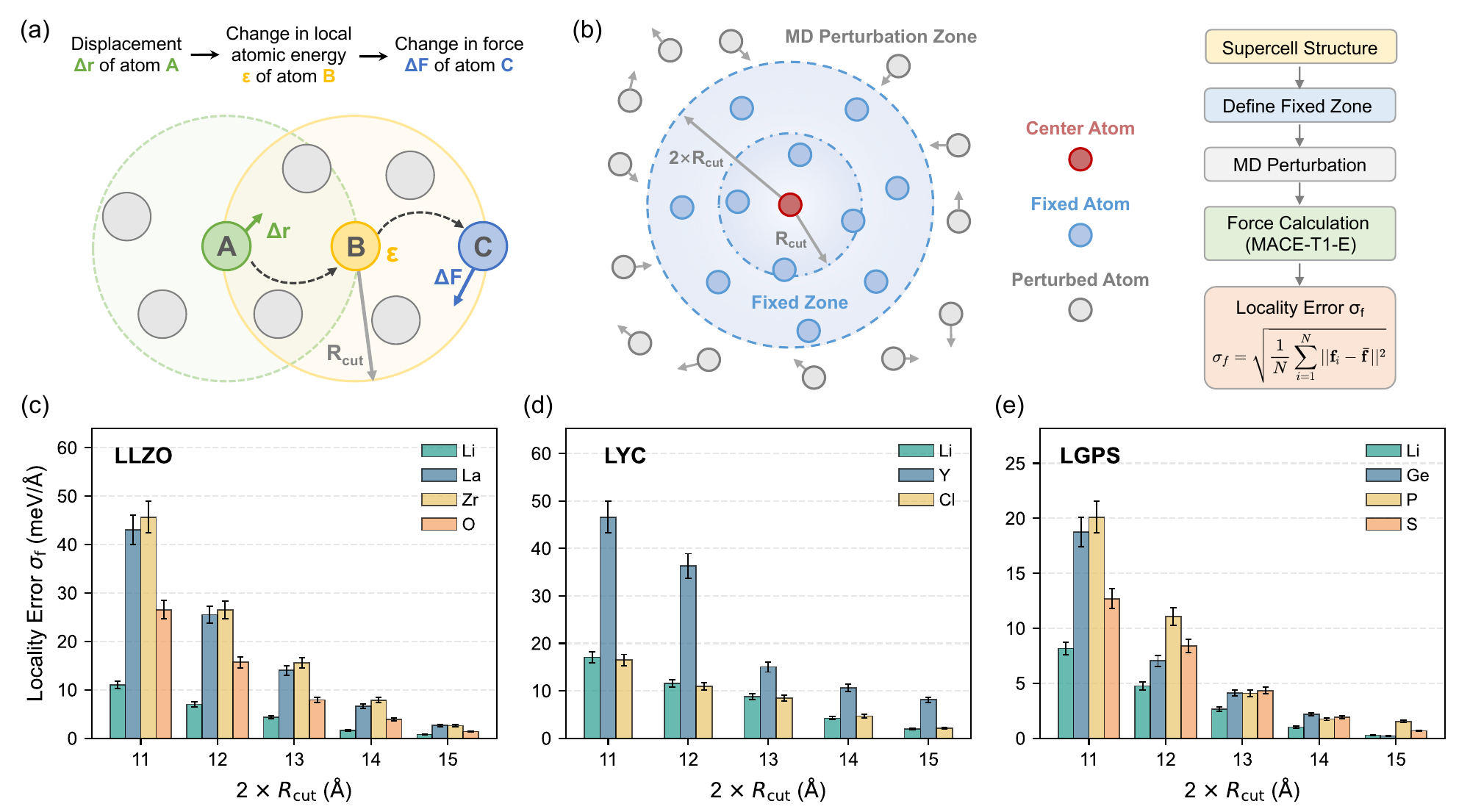}
	\caption{(a) Schematic of force locality in atom-centered many-body potentials (adapted from Ref~\cite{deringer2021gaussian}): the displacement of atom A influences the force on atom C via their shared neighbor B, extending the force locality to twice the cutoff radius ($2\times R_{cut}$). (b) Schematic of the locality test: atoms within $2\times R_{cut}$ of the central atom are fixed, while atoms in the outer region are subjected to MD perturbations. (c–e) Locality errors versus $2\times R_{cut}$ for LLZO, LYC, and LGPS, respectively.}
	\label{fig:force_dev}
\end{figure*}

Most SSEs are ceramic-based ionic materials. Conventionally, long-range Coulomb interactions are considered essential for describing such systems. As a result, empirical force fields for SSEs typically incorporate an explicit long-range Coulomb term~\cite{klenk2016finite}. However, MLFFs handle the atomic interactions differently depending on their architectures, which may employ strictly local descriptors, schemes for message passing, or explicit long-range corrections.

Consequently, it remains unclear to what extent these long range Coulomb interactions actually influence the interatomic forces in SSEs, or whether the crystal environment effectively screens them. Building on this, our previous findings raise a compelling question: if long-range interactions are physically critical, why are the predictions of the ionic transport behavior from the strictly local NEP model almost identical to those of the non local MACE models? This paradox suggests a need to investigate how localized the interactions between atoms in SSEs truly are and whether explicit terms for long-range effects are indeed dispensable for reliable ionic transport simulations. To address this, we conducted systematic locality tests to quantify the effective spatial range of force interactions in these materials.

The use of finite cutoff radius ($R_{cut}$) in MLFFs imposes an inherent limitation on the range of interactions they can capture. For atom-centered descriptors involving many-body terms, the effective range of force interactions extends to $2 \times R_{cut}$~\cite{deringer2021gaussian}, as illustrated in Figure~\ref{fig:force_dev}(a). When atom A is displaced, it perturbs the local atomistic energy of atom B within $R_{cut}$. This energy change then affects the force on atom C, which also lies within $R_{cut}$ of atom B. Consequently, atoms A and C can be separated by up to $2\times R_{cut}$ while still influencing the forces of each other.

Meanwhile, this locality means that any genuine long-range interactions beyond $2\times R_{cut}$ in the quantum-mechanical description will appear as noise in the training data, since any atomic configurations outside of $2\times R_{cut}$ are not taken into consideration by the short-range MLFFs. For a given material, understanding the magnitude of this locality error is crucial, as it defines the lower bound of achievable force error for any short-range MLFF with a specified $R_{cut}$. This lower bound error estimation is independent of the MLFF architecture. To quantify this intrinsic limitation for MLFFs, we performed systematic locality tests on SSE materials.

Our testing protocol is illustrated in Figure~\ref{fig:force_dev}(b). For a selected central atom, we defined two concentric zones: an inner fixed zone of radius $2\times R_{cut}$ where all atomic positions remain frozen, and an outer perturbation zone where atoms undergo MD simulations. Unlike previous studies that employed manual random displacements~\cite{huang2021deep,staacke2021role}, we chose MD-based perturbations because the magnitude of manual displacements can significantly affect the measured locality error, as noted in earlier work~\cite{deringer2021gaussian}. MD perturbations naturally sample the Boltzmann distribution at the target temperature, providing a more physically reasonable configurations. To ensure sufficient atomic displacements, we applied MD perturbations at 1000 K to LLZO and 500 K to LYC and LGPS.

The specific implementation for LLZO exemplifies our approach (Figure~\ref{fig:force_dev}(b)). We constructed a $3\times3\times3$ supercell containing 5,184 atoms in a cubic box with a length of 39.3 Å. Central atoms of each species (Li, La, Zr, O) were selected near the center of box. For each central atom, we performed 500 ps NVT simulations at 1000 K using GPUMD, extracting 100 equally spaced configurations. Since these large supercells far exceed practical DFT computational capacity, we employed MACE-T1-E model as the our DFT surrogate, which with its message passing architecture can capture interactions extending to approximately $4\times R_{cut}$, well beyond the $2\times R_{cut}$ locality limit.

To quantify force deviations, the locality error ($\sigma_{f}$) is defined as the magnitude of the force fluctuations on the central atom induced by outside perturbations beyond the $2\times R_{\text{cut}}$. This metric is quantified by calculating the standard deviation of the atomic force vector $\mathbf{f}$ across all sampled configurations:

\begin{equation*}
\sigma_{f} = \sqrt{\frac{1}{N} \sum_{i=1}^{N} || \mathbf{f}_i - \bar{\mathbf{f}} ||^2},
\end{equation*}
where $\bar{\mathbf{f}} = \frac{1}{N} \sum \mathbf{f}_i$ is the average force vector and $\|\cdot\|$ denotes the Euclidean norm.  This locality error provides a direct measure of the sensitivity of atomic forces to long-range structural variations.

We found that force deviations decay rapidly with increasing $2\times R_{\text{cut}}$ across all three materials (Figure~\ref{fig:force_dev}(c-e)). As expected, the magnitude of $\sigma_{f}$ correlates strongly with ionic charge states. High-valence cations such as La$^{3+}$ and Zr$^{4+}$ exhibit significantly larger force deviations than Li$^{+}$, while O$^{2-}$ shows intermediate values exceeding those of Li$^{+}$. Similar trend also exhibits in LYC, where the $\sigma_{f}$ for Y$^{3+}$ is approximately three times larger than for monovalent Li$^{+}$ and Cl$^{-}$, directly reflecting the threefold difference in formal charges. In LGPS, the higher formal charges (Ge$^{4+}$ and P$^{5+}$) are also associated with larger force deviations.

When the $R_{\text{cut}}$ is approximately 6 Å, $\sigma_{f}$ becomes very small. Only the Y$^{3+}$ ion in LYC maintains a significant deviation at this cutoff radius, which may explain why the previous study on LYC used a $R_{cut}$ of 7.5 Å~\cite{wang2023frustration}. The limited force deviations for Li-ions carry important physical implications: Li-ion diffusion is predominantly governed by local atomic environments rather than long-range interactions. Therefore, short-range MLFFs, such as NEP and MACE-T0 models, despite lacking explicit long-range electrostatics or message passing mechanisms, can still produce diffusivity predictions comparable to those advanced models.

These findings suggest that long-range electrostatic interactions may play a more modest role in bulk SSE than often assumed. With sufficiently large cutoffs, short-range models can capture Li-ion diffusion mechanisms without explicit charge treatments. However, this conclusion applies specifically to defect-free bulk crystals. Long-range and charge-explicit models are expected to be critical for certain systems, such as grain boundaries, surfaces, and defect environments where charge redistribution and electrostatic fields may extend over longer distances. Moreover, the explicit prediction of environment-dependent partial charges offers a novel analytical dimension. For instance, recent work utilizing the qNEP framework demonstrated that monitoring these local charge distributions provides a unique way for characterizing the order-disorder phase transition in LLZO and tracking charge evolution pathways at magnesium-water interfaces~\cite{fan2026qnep}.

\section{Summary and Outlook}

In this perspective, we systematically evaluated a few key factors for developing MLFFs for SSEs. Our benchmarks reveal that the PESs of bulk SSEs are relatively simple. Due to their rigid frameworks and constrained diffusion paths, effective force fields can be trained using modest-size datasets. Crucially, we find that force training error is not a reliable indicator to evaluate MLFF performance on ion transport properties, and data quality matters much more than quantity.

Looking forward, although our results suggest that explicit long-range Coulombic interactions have a limited impact on bulk diffusion in SSEs, the treatment of charge remains essential for systems that violate charge neutrality. In regions such as grain boundaries, charged defect centers, and electrode interfaces, non-uniform distribution of charge can significantly influence local electric fields and ionic dynamics. Furthermore, the simulation of complex interfaces presents additional challenges, particularly regarding lithium dendrite growth and redox reactions at the electrodes. These processes require force fields with reactive capabilities, which can dynamically infer atomic charges based on the evolving local chemical environment. These processes demand architectures that can dynamically infer atomic charges based on the evolving chemical environment. Consequently, further development of MLFFs to dynamically infer atomic charges based on the evolving chemical environment, and to accurately incorporate long-range interactions will be vital for modeling these complex cases.

Finally, while this work has focused on specialized force fields trained for specific systems, the emergence of foundation models offers a promising new paradigm. Although these models may be less accurate compared to customized potentials, their ability to perform simulations without training from scratch makes them highly valuable for screening vast chemical spaces. In the future, the integration of high-precision specialized models with broad-scale foundation models will likely provide a multi-scale computational framework for the rational design of next-generation solid-state battery materials.

\section*{Data and Software Availability}
After the perspective is published, the source files will be available in our GitHub repository at \href{https://github.com/zhyan0603/SourceFiles}{\textcolor{DarkBlue}{https://github.com/zhyan0603/SourceFiles}}. The source code and tutorials for our GPUMDkit are available at \href{https://github.com/zhyan0603/GPUMDkit}{\textcolor{DarkBlue}{https://github.com/zhyan0603/GPUMDkit}}.

\section*{Declaration of competing interest}
The authors have no conflicts to disclose.

\section*{Contributions}
Y. Zhu and Z. Yan conceived and designed the research; Y. Zhu guided the research; Z. Yan performed the simulations; Z. Yan, S. Tang, and Y. Zhu wrote the paper.

\section*{Acknowledgements}
We acknowledge the support from the National Natural Science Foundation of China (No. 22509162 and 225B2917). The computational resource is provided by the Open Source Supercomputing Center of S-A-I.

\section*{Supporting information}
The Supporting Information will be available free of charge at npj Energy Materials website.

\bibliography{references}
\end{document}